\documentclass[12pt]{iopart}
    \input epsf
    \begin{document}

  \title[Hitting a BEC with a comb]{Hitting a BEC with a comb:
Evolution of interference
patterns inside a magnetic trap}
      \author{J.H.~M\"uller, O.~Morsch, M.~Cristiani, D.~Ciampini, and
E.~Arimondo}
      \address{INFM, Dipartimento di Fisica E.Fermi, Universit\`{a} di Pisa, Via
      Buonarroti 2, I-56127 Pisa,Italy}
      \date{\today}
      \begin{abstract}
      We study the evolution inside a harmonic trap of
      Bose-Einstein condensates released from the periodic potential
      of an optical lattice. After a time-of-flight, harmonic motion
      of the interference peaks is observed as well as a breathing
      motion in the direction perpendicular to the optical lattice.
      We interpret these results in terms of a simple physical model
      and discuss the possibility of more detailed studies of such a
      system.
      \end{abstract}
      \maketitle
      \section{Introduction}
      Optical lattices have been used in a number of experiments on
      Bose-Einstein condensates (BECs) in recent years. The periodic
      potential provided by these lattices has been exploited to
      study, amongst other things, dynamical phenomena such as Bloch
      oscillations~\cite{anderson98,morsch01,cristiani02}, the
      Bogoliubov dispersion relation~\cite{stamperkurn99,steinhauer02}
and Josephson
      oscillations~\cite{cataliotti01} as well as phase properties as
      manifested in number squeezing~\cite{orzel01}, the Mott
      insulator transition~\cite{greiner02} and coherent state
      revivals~\cite{greiner02a}. In most of these experiments, the
      interference pattern from the condensates trapped at the
      potential minima of the lattice was studied by switching off
      the optical lattice and magnetic trap simultaneously.

      In~\cite{morsch02}, we have already demonstrated that an
      experimental protocol in which only the magnetic trap is
      switched off and the condensate is then allowed to expand
      freely inside the (one-dimensional) optical lattice can be used
      to infer the chemical potential of the condensate in the
      combined harmonic plus periodic potential. In the present work,
      we use a variation on this theme: instead of switching off the
      magnetic trap and leaving the optical lattice on, we suddenly
      switch off the optical lattice and allow the condensate to
      evolve inside the magnetic trap for a variable time.
      Thereafter, the magnetic trap is also switched off and the
      condensate is observed after a time-of-flight.

      By suddenly switching off the optical lattice, we
      instantaneously release the condensate from a strongly deformed
      potential created locally by the individual wells of the
      optical lattice. In this way, one expects to excite the full
      spectrum of possible shape oscillations along the lattice
      direction and, through the non-linear coupling between the
      degrees of freedom, also in the other directions.

      After the conclusion of our experiments, a theoretical paper
      discussing a scheme similar to ours was brought to our
      attention~\cite{xiong02}, and more recently a generalization
      to three dimensions was explored in~\cite{adikhari02}.

      This paper is organized as follows. In section~\ref{setup}, we
      outline our setup and the experimental protocol used.
      Section~\ref{results} presents the experimental
      results, followed by a discussion (Section~\ref{discussion}).
Finally, we present our
      conclusions and an outlook on further
      experiments in section~\ref{conclusions}.

      \section{Experimental setup and protocol}\label{setup}
      In a triaxial time-orbiting potential (TOP) trap with magnetic
      trapping frequencies in the ratio $\omega_x:\omega_y:\omega_z$
      of $2:1:\sqrt{2-\epsilon}$ ($\epsilon$ goes to zero for large
trapping frequencies and is typically of the order of $0.2$
      in the present experiment; $z$ is the downward direction), we
create BECs of around $1.5\times 10^4$
      rubidium atoms as described in detail in~\cite{muller00}. Once
      the BEC has formed, the TOP trap is adiabatically relaxed from
      a mean trap frequency $\overline{\omega}$ of about
      $100\,\mathrm{Hz}$ to $\approx 30\,\mathrm{Hz}$. Two linearly
      polarized laser beams with parallel polarizations intersecting
      at a (full) angle of $38$ degrees at the position of the BEC
provide a sinusoidal optical lattice potential
       of the form $V(z)=V_0\sin^2(\pi z/d)$ with
      period $d=1.2\,\mathrm{\mu m}$. This lattice is superimposed onto
the harmonic
      potential of the magnetic trap. The detuning of the lattice
      beams from the atomic resonance is $\approx 30\,\mathrm{GHz}$
      and the intensity around $10\,\mathrm{mW\,cm^{-2}}$ (for a beam
      waist of $1.8\,\mathrm{mm}$), leading to a lattice depth $V_0$ of
      about $15\,E_{rec}$, measured in units of the lattice recoil
      energy $E_{rec}=\hbar^2 \pi^2/(2md^2)$ (where $m$ is the mass
      of the rubidium atoms).

\begin{figure}
      \centering\begin{center}\mbox{\epsfxsize 2.65 in \epsfbox{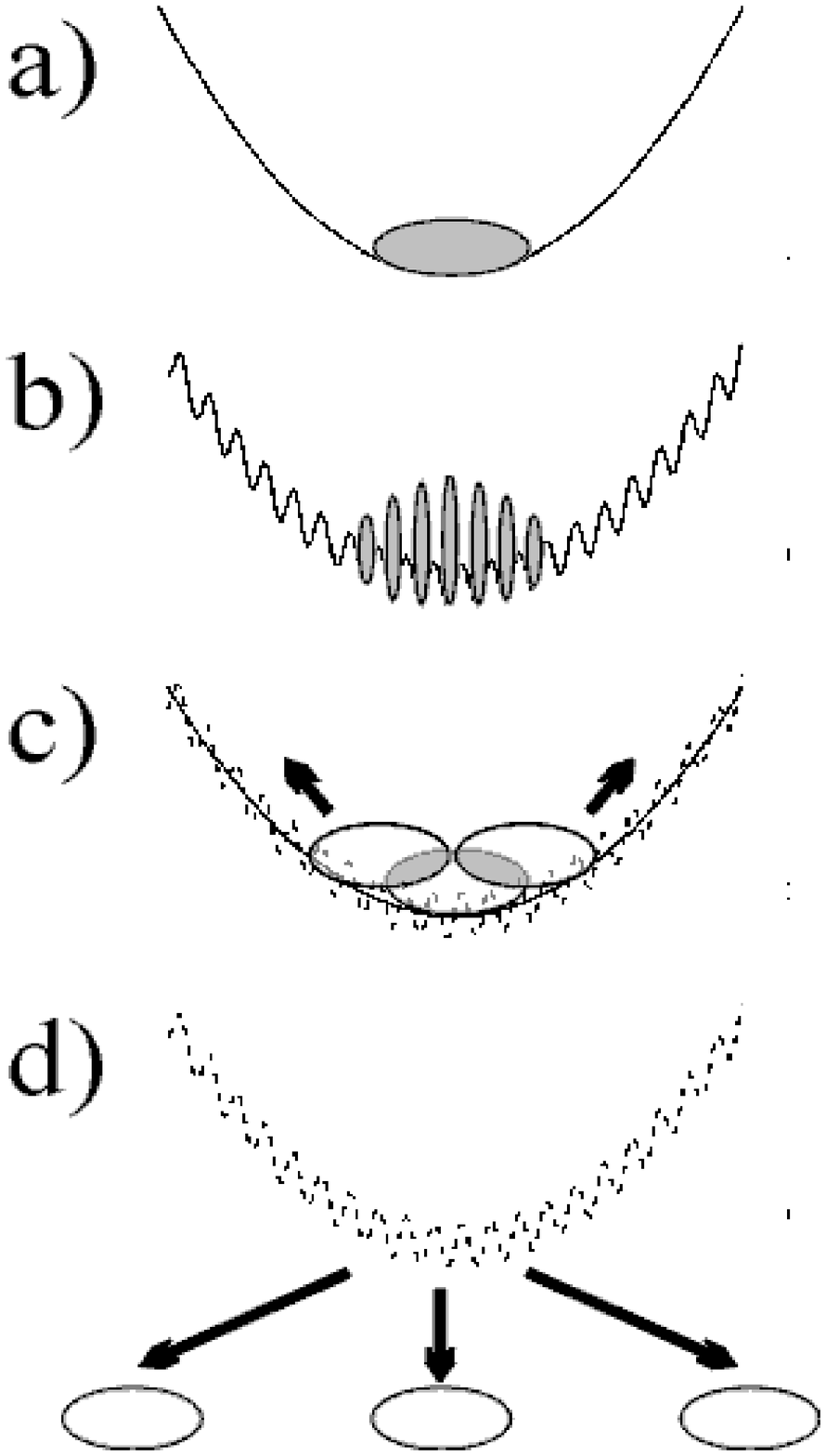}}
      \caption{Experimental protocol. After creating a BEC in the magnetic
      trap (a), the optical lattice is switched on (b) for $1\,\mathrm{ms}$.
      When the lattice is switched off again (c), the different momentum classes
      created by the lattice evolve freely inside the harmonic potential.
      This evolution is then monitored by switching off the magnetic trap
      and observing the BEC after a time-of flight (d).}\label{Fig_1}
      \end{center}\end{figure}

      In an experimental cycle, after creating the BEC the optical
      lattice is switched on with a linear ramp of duration
      $\tau=1\,\mathrm{ms}$ and then left at the final lattice depth
      for $1\,\mathrm{ms}$. Thereafter, it is suddenly switched off
      and the BEC is allowed to evolve inside the magnetic trap for a
      duration $t$. The switch-on time is chosen such that, in the
      Bloch band picture~\cite{cristiani02}, the BEC remains in the
      ground state of the band structure. With respect to the magnetic
      trap frequencies $\omega_i$, however, the switching-on procedure
is non-adiabatic as $1/\tau\gg \omega_i$.
      In this way, we "hit" the BEC with the comb-like lattice
potential. After the evolution time
      $t$, the magnetic trap is also switched off and the expanded
      BEC is observed after a time-of-flight of $21\,\mathrm{ms}$.
      The experimental protocol is illustrated in Fig.~\ref{Fig_1}.

      \section{Experimental results}\label{results}
      When a BEC is released from an optical lattice (in the absence
      of a
      magnetic trap), after a
      time-of-flight $t_{TOF}$ the periodic spatial pattern of the
      condensate density produces an interference pattern that is
composed of a central peak
      and side-peaks at positions $z_{side}=\pm 2nv_{rec}t_{TOF}$, where
      $v_{rec}=h/(2md)$ is the lattice recoil velocity and $n$ is an
      integer~\cite{anderson98,morsch01,cataliotti01}. For modest
lattice depths ($<20\,E_{rec})$ only the
      side-peaks corresponding to $n=1$ will be visible as the
      higher-order peaks are suppressed by the finite width of the
      condensates expanding from the individual wells. The population of
      the side-peaks relative to that of the central peak can be used
      to infer the widths of the locally trapped condensates and hence
the lattice depth~\cite{cristiani02}.

\begin{figure}
      \centering\begin{center}\mbox{\epsfxsize 2.65 in \epsfbox{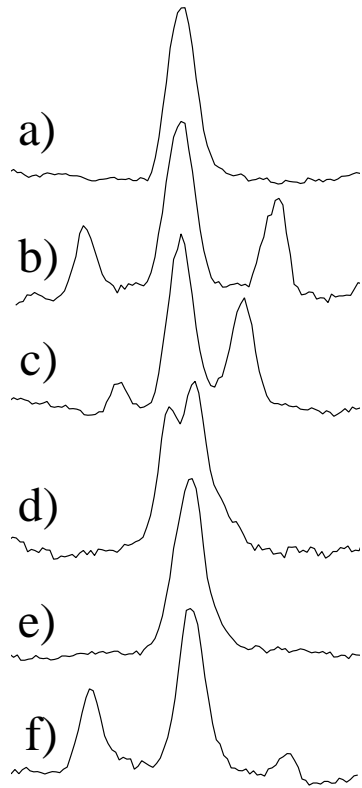}}
      \caption{Evolution of the interference pattern (integrated perpendicular
      to the lattice direction) of a condensate released from the magnetic trap
      without (a) and after applying an optical lattice ((b)-(f)) for
      $1\,\mathrm{ms}$. The evolution times inside the trap for (b)-(f) are
      $0.1$, $6$, $8$, $10$ and $18\,\mathrm{ms}$, respectively.}\label{Fig_2}
      \end{center}\end{figure}

\begin{figure}
      \centering\begin{center}\mbox{\epsfxsize 3.5 in \epsfbox{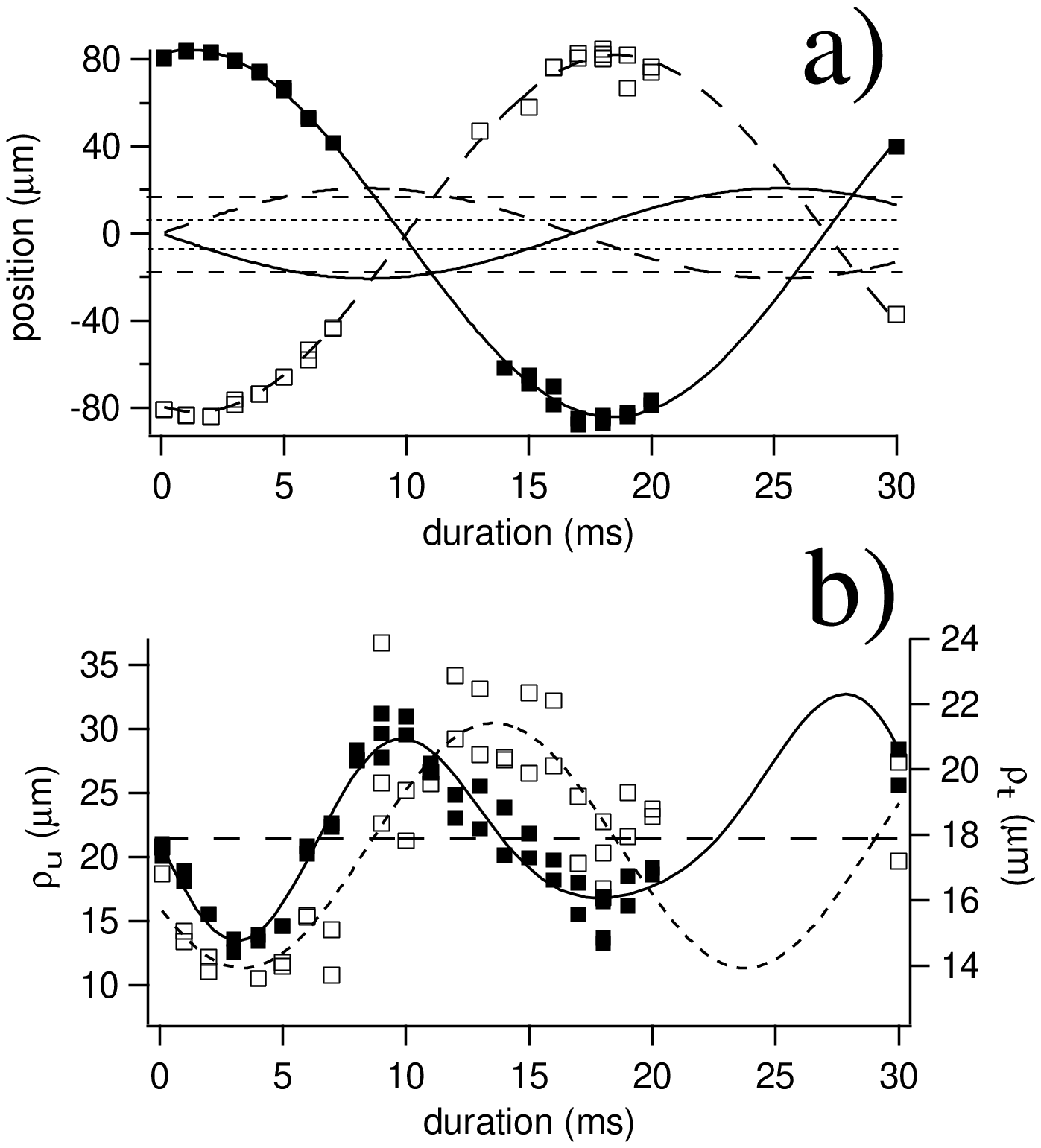}}
      \caption{Condensate evolution inside the magnetic trap after suddenly
      switching off the optical lattice. The harmonic motion of the sidepeaks
      is visible in (a). The solid and dashed sinusoidal curves with the smaller
      amplitude represent the positions of the sidepeaks inside the trap as
      calculated from the parameters extracted from the fit to the experimental
      points after a time-of-flight. The horizontal dotted and dashed lines
      indicate the width of the condensate inside the trap and after a
      time-of-flight of $21\,\mathrm{ms}$, respectively. In (b), the widths
      perpendicular to (filled symbols) and along (open symbols) the lattice
      direction are shown along with the value measured without the optical
      lattice (dashed horizontal line). The solid and dashed sinusoidal curves
       are sine-fits to the experimental points; in (b), the solid line is a
       fit using three frequencies (see text).}\label{Fig_3}
      \end{center}\end{figure}

      For the present experimental investigation with release from the
      optical lattice into the magnetic trap, the periodic spatial
      pattern created by the optical lattice produces the interference
      pattern described above, but that pattern is modified by the
      magnetic trap evolution. Figure~\ref{Fig_2} shows the
interference pattern (monitored
      after the time of flight and integrated
      along the direction perpendicular to the lattice direction) for
      various evolution times inside the magnetic trap after the
lattice is switched off.
      We observe two side-peaks whose positions vary in time as the
      released condensate evolves inside the magnetic trap. This
      behaviour is also evident in Fig.~\ref{Fig_3} (a). Here, we
      have plotted the position of the two side-peaks~\cite{footnote_sidep}
      relative to the central peak (in order to
      account for residual sloshing of the condensate) as a function
      of $t$. The unequal heights of the two side-peaks in Fig.~\ref{Fig_2}
      are due to sloshing (and hence a non-zero initial velocity relative
      to the lattice) of the condensate
      inside the magnetic trap before the lattice is switched on.

      In Fig.~\ref{Fig_3} (b), the widths $\rho_{||}$ and $\rho_\perp$
of the central peak along and perpendicular to the lattice
      direction, respectively, are shown as a function of time. One
      clearly sees oscillations of both the radial and the
      longitudinal widths.

      \section{Discussion}\label{discussion}
       An optical lattice creates a regular spatial pattern of fragmented
      condensates, each of them being trapped by the local mimimum of the
      lattice potential. When the optical lattice is switched off, the
matter waves emitted from
      each individual condensate interfere. In the far field, the
emission from a
      one-dimensional array of coherent sources produces
      a regular diffraction pattern with a central maximum and
      symmetric, equally spaced side-peaks. The separation between the
      central peak and the side-peaks is determined by the periodic spatial
      pattern  of the condensate. The periodic evolution inside the magnetic
      trap of the condensate spatial pattern initially created by the
      optical lattice produces the observed oscillation of the side peaks.

      The theoretical analysis of ref.\cite{xiong02} provides a simple
      picture for the interpretation oscillatory
      motion of the side-peaks. When the optical lattice is removed,
the interference of
      the condensate spatial pattern with initial
      period $d$  inside the magnetic trap produces a periodic pattern
with a central peak located
      at $x=0$ and two side peaks. This fragmented structure of the
      condensate experiences the confinement of the
      harmonic magnetic trap potential. As a consequence, the two
      side-peaks execute an oscillatory motion. The three-peak structure
      experimentally observed after the time of flight is produced by
      the three-peak fragmentation of the condensate taking place
      inside the magnetic trap when the optical lattice is removed.

      The observed motion of the two side-peaks
      can be explained by considering independently the evolution inside
      the magnetic trap of the three interference peaks created after
switching off the
      optical lattice~\cite{xiong02}.
Since the lattice transfers momentum to the condensate in units of
$2p_{rec}=2mv_{rec}$,
      the side-peaks initially move with velocities
      $\pm 2v_{rec}$ at $t=0$ and then
      individually perform a harmonic motion of the form
      \begin{equation}
      z(t)=\pm \frac{h}{m\omega_z d}\sin(\omega_z t),
      \label{sin}
  \end{equation}
      from which their expected positions after a time-of-flight can
      be easily calculated. The intuitive picture of the
      side-peaks moving like individual BECs is only valid once the
      condensates at the lattice sites have expanded sufficiently
      to overlap with each other and hence produce the three-peak interference
      pattern. For our experimental parameters, this time is less
      than $0.1\,\mathrm{ms}$. In Fig.~\ref{Fig_3} (a), we have fitted
      the sinusoidal function of Eq. \ref{sin} for the expected position to the
      experimental points. The fitted values for $\omega_z$ and $d$
      agree to within the experimental error with values obtained
      independently from measurements of the dipole (sloshing)
      frequency of a single condensate and the lattice spacing quoted
      above, respectively. The oscillation amplitude $h/(m\omega_z
      d)\approx 20\,\mathrm{\mu m}$ is about twice as large as the
      Thomas-Fermi radius of the condensate inside the magnetic trap
      for the parameters of the present experiment.

      Apart from the appearance of interference peaks along the
      lattice directions, one also expects to see shape oscillations
      of the condensate in the radial directions, as during the
      lattice phase the condensate gets broken up into individual
      condensates localized at the lattice sites (for $V_0=15\,E_{rec}$,
the width of a wave-packet
      inside a lattice well is $\approx 0.17 d$). When the lattice is
      suddenly switched off, the effective trapping frequency in the
      lattice direction jumps from $\approx 2600\,\mathrm{Hz}$ (the
      harmonic frequency at a lattice site) to $30\,\mathrm{Hz}$.
      The condensate will, therefore, start expanding along the
      lattice direction and subsequently perform breathing
      oscillations inside the magnetic trap. As the degrees of
      freedom of the condensate are coupled through the non-linear
      interaction term in the Gross-Pitaevskii equation, the
      condensate will start breathing in the radial directions as
      well.

      In practice, we expect to find frequency components
      corresponding to the breathing motions in the three directions
      of the trap in the evolution of all three in-trap
widths~\cite{footnote_sim}. After
      releasing the condensate from the trap, the in-trap width of the
central interference peak will
      still be directly reflected in the radial direction (after taking
into account the expansion), whereas
      the width in the lattice direction is not so easy to interpret
      because of interference effects. Fig.~\ref{Fig_3} shows a fit
      to $\rho_\perp$ using a fitting function with three sinusoidal
      oscillations with frequencies in the ratio of the three
      harmonic frequencies of our TOP-trap and independent
      amplitudes. This fit gives a frequency of $59.5\,\mathrm{Hz}$
      for the dominant contribution, corresponding to $\approx
      2\times\omega_z/(2\pi)$. A simple sinusoidal fit to the
      $\rho_\perp$-data gives a similar result for the frequency, but
      fits the experimental points considerably less well. The radial
      widths of the two side-peaks oscillate with the same frequency
      as the central peak (and with the same phase).

      For the longitudinal width $\rho_{||}$, we have used a simple
      sinusoidal fit, giving a frequency of
      $49\,\mathrm{Hz}\approx1.6\times \omega_z/(2\pi)$. As mentioned
      above, $\rho_{||}$ does not simply reflect the in-trap
      longitudinal width of the individual condensates but is
      determined by the interference between all of them.

      \section{Conclusion and outlook}\label{conclusions}
      In this work, we have investigated the evolution inside a
      magnetic trap of a condensate "hit" by a periodic potential.
      The condensate evolution inside the magnetic trap produces a three
      peak fragmented structure of the condensate. Those peaks perform
      an oscillating motion at the magnetic trap
      frequency along the lattice direction, as viewed after the
      time-of-flight.   Apart from the appearance of side-peaks, one
can also study the
      breathing motion of the central interference peak. In this
      sense, the present work ties in with the investigations of
      collective modes of BECs~\cite{mewes96,jin96}. So far, both
experimental and
      theoretical results have been obtained for low-lying modes of a
      condensate in a magnetic trap, including breathing modes~\cite{chevy02},
      surface modes~\cite{onofrio00} and the scissors
mode~\cite{marago00}. As in the present work,
      in these experiments the collective modes were typically excited by a
      sudden change in the trap frequency or geometry, and the frequency and
      damping rate of the subsequent oscillations were measured either
      in situ or after a time-of-flight.

      In ref.~\cite{xiong02}, the effect of interactions on the evolution
      of the various peaks moving {\em inside} the trap was
      discussed.  The mean field interaction should produce a deviation
      for the motion of the side-peaks from the simple oscillating
      motion at frequency $\omega_{z}$. It would be interesting to investigate
      the effect of the mean field on the widths of the peaks after
      a time-of-flight and to compare these results to the
      non-interacting case. Furthermore, it should be investigated
      whether
      the mean field interaction could produce soliton-like
      features
      with the spatial pattern of the condensate and preserving its shape.

      In future experiments, our method could be used to study in
      more detail the frequency spectrum of the radial breathing
      oscillations, requiring data for several oscillation cycles to
      give a reasonable resolution, as well as possible damping
      mechanisms both for the radial oscillations and the
      oscillations of the side-peaks. In our experiments, for
      $t>15\,\mathrm{ms}$, corresponding to the first "collision" of
      the side-peaks inside the trap, we see collisional haloes
      which hint at a possible damping mechanism.

      \section*{Acknowledgments}
      This work was supported by the MURST (PRIN2000
      Initiative), the INFM (Progetto di Ricerca Avanzata
      `Photonmatter'), and by  the EU through the Cold Quantum
      Gases Network, Contract No. HPRN-CT-2000-00125. O.M. gratefully
      acknowledges financial support from the EU within the IHP
      Programme.

       \newpage

\begin{thebibliography}{99}
\bibitem{footnote_sidep}After the first "crossing" at $t\approx
      10\,\mathrm{ms}$, we have swapped the labels "left" and "right"
      for the side-peaks to account for the intuitive picture of the
      "right" sidepeak appearing to the left of the central peak
      (after a time-of-flight) after inverting its velocity inside
      the trap.
\end{thebibliography}
    \end{document}